\documentclass[twocolumn, amsmath,amssymb, nobibnotes, aps, prb, superscriptaddress, citeautoscript]{revtex4-2}

\usepackage[colorlinks=true,linkcolor=blue,citecolor=blue,urlcolor=blue]{hyperref}
\usepackage[T1]{fontenc}
\DeclareUnicodeCharacter{00A0}{ }
\usepackage{graphicx}
\usepackage{gensymb}
\usepackage{amsmath,bm}
\usepackage{braket}
\usepackage[normalem]{ulem}
\usepackage[shortlabels]{enumitem}
\usepackage[labelfont=bf]{caption}
\usepackage{titlesec}
\usepackage{setspace}

\usepackage{dcolumn}
\usepackage{subfigure}
\usepackage{natbib}
\usepackage{multirow}
\usepackage{soul}
\usepackage{float}
\usepackage[normalem]{ulem}
\usepackage[usenames,dvipsnames]{color}
\usepackage{appendix}
\usepackage{mathrsfs}
\usepackage[colorlinks=true,linkcolor=blue,citecolor=blue,urlcolor=blue]{hyperref}
\usepackage{mathrsfs}

\begin{document}

\title{Unraveling the effects of inter-site Hubbard interactions in spinel Li-ion cathode materials}

\author{Iurii Timrov}\email[ ]{ iurii.timrov@epfl.ch}
\affiliation{Theory and Simulation of Materials (THEOS), and National Centre for Computational Design and Discovery of Novel Materials (MARVEL), \'Ecole Polytechnique F\'ed\'erale de Lausanne (EPFL), CH-1015 Lausanne, Switzerland}

\author{Michele Kotiuga}
\affiliation{Theory and Simulation of Materials (THEOS), and National Centre for Computational Design and Discovery of Novel Materials (MARVEL), \'Ecole Polytechnique F\'ed\'erale de Lausanne (EPFL), CH-1015 Lausanne, Switzerland}

\author{Nicola Marzari}
\affiliation{Theory and Simulation of Materials (THEOS), and National Centre for Computational Design and Discovery of Novel Materials (MARVEL), \'Ecole Polytechnique F\'ed\'erale de Lausanne (EPFL), CH-1015 Lausanne, Switzerland}

\begin{abstract}
Accurate first-principles predictions of the structural, electronic, magnetic, and electrochemical properties of cathode materials can be key in the design of novel efficient Li-ion batteries. Spinel-type cathode materials Li$_x$Mn$_2$O$_4$ and Li$_x$Mn$_{1.5}$Ni$_{0.5}$O$_4$ are promising candidates for Li-ion battery technologies, but they present serious challenges when it comes to their first-principles modeling. Here, we use density-functional theory with extended Hubbard functionals---DFT+$U$+$V$ with on-site $U$ and inter-site $V$ Hubbard interactions---to study the properties of these transition-metal oxides. The Hubbard parameters are computed from first-principles using density-functional perturbation theory. We show that while $U$ is crucial to obtain the right trends in properties of these materials, $V$ is essential for a quantitative description of the structural and electronic properties, as well as the Li-intercalation voltages. This work paves the way for reliable first-principles studies of other families of cathode materials without relying on empirical fitting or calibration procedures.
\end{abstract}

\date{\today}

\maketitle

\section{Introduction}
\label{sec:intro}

Over the past 30 years rechargeable Li-ion batteries have been the subject of very active research, driven both by the growing worldwide market of portable electronics and electric vehicles and in order to meet the climate-neutral requirements of our society~\cite{Tarascon:2014, Kang:2009}. Applications demand advanced energy-storage systems with excellent energy density, cyclability, and thermal stability. Among various available cathode materials, lithium-manganese-oxide spinels have attracted special attention due to their low cost, nontoxicity, and higher Li-intercalation voltages than commercial layered rock-salt cathodes~\cite{Thackeray:1983}.

LiMn$_2$O$_4$ is the prototypical spinel cathode material~\cite{Huang:2021}, having an operating voltage of $\sim 4.15$~V~\cite{Ohzuku:1994, Barker:1995, Zhou:2004b} and a high capacity of $\sim 140$~mA h g$^{-1}$~\cite{Kanno:1999}. This material has a high-temperature cubic phase with space group $Fd\bar{3}m$; however, at low temperatures it is still debated whether the structure is orthorhombic or tetragonal~\cite{Yamada:1995, RodriguezCarvajal:1998, Huang:2011}. The Mn ions have mixed-valence state consisting of Jahn-Teller (JT) active Mn$^{3+}$ and non-JT active Mn$^{4+}$ ions, and, while theses are disordered in the cubic phase, in the low-temperature phase there is charge ordering of these two types of Mn ions~\cite{RodriguezCarvajal:1998}. Experimentally it is know that LiMn$_2$O$_4$ has an antiferromagnetic (AFM) ground state at low temperatures~\cite{Wills:1999}, though the exact type of AFM ordering is not known due to the complex magnetic interactions in this material. However, despite the attractive properties of LiMn$_2$O$_4$ as a cathode material, it exhibits degradation with extended cycling, and hence various types of doping (partial Mn substitutions) have been explored in order to enhance the electrochemical performance~\cite{Zhong:2012, Molenda:2004, MoorheadRosenberg:2013}. One candidate that has received a great deal of attention is LiMn$_{1.5}$Ni$_{0.5}$O$_4$, as it offers even higher power capabilities with an operating voltage of $\sim 4.7$~V and a capacity of $\sim 135$~mA h g$^{-1}$~\cite{Muraliganth:2010, Manthiram:2014}. It is known experimentally that LiMn$_{1.5}$Ni$_{0.5}$O$_4$ also crystallizes in a spinel phase but in two possible space groups $Fd\bar{3}m$ or $P4_332$ (or $P4_132$, which is an enantiomorph of $P4_332$) depending on the synthesis conditions, and that Ni$^{2+}$ is redox active while Mn$^{4+}$ is inactive~\cite{Manthiram:2014, Amin:2020, Stuble:2023}. This material adopts a ferrimagnetic (FiM) ordering below the Curie temperature ($\sim 130$~K) in which the Mn$^{4+}$ spins align antiparallel to the Ni$^{2+}$ spins~\cite{Amdouni:2007, MoorheadRosenberg:2012, MoorheadRosenberg:2013}. Nonetheless, the commercialization of this material has been hampered by severe capacity fade, particularly at elevated temperatures, in cells employing a graphite anode~\cite{Manthiram:2014}. Therefore, the need is apparent for in-depth characterization of these materials in order to understand the underlying physical mechanisms and to further improve their electrochemical performance.

\begin{figure*}[t]
  \centering
  \includegraphics[width=0.85\linewidth]{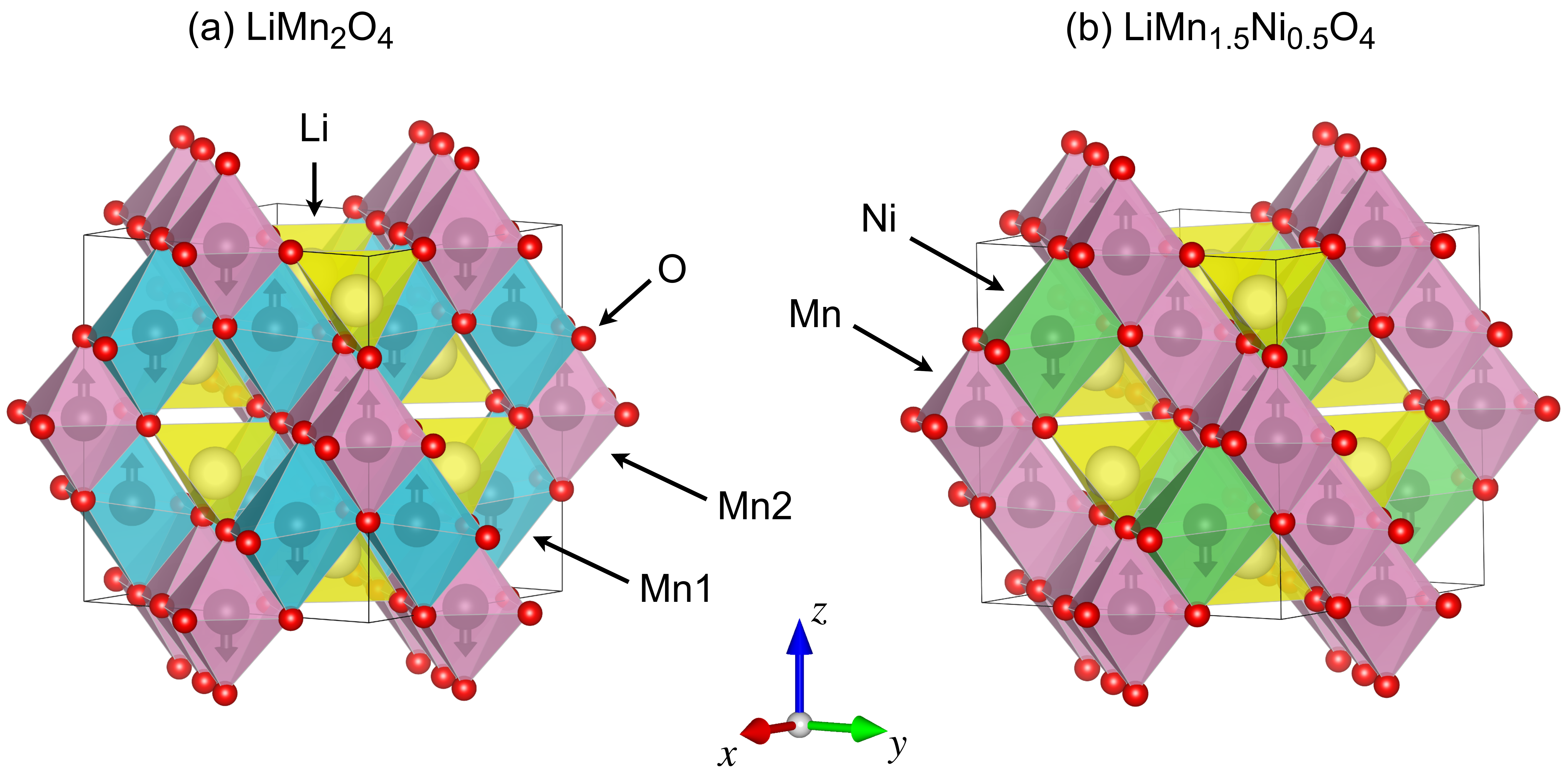}
   \caption{Crystal structure of spinel cathode materials: (a)~LiMn$_2$O$_4$, and (b)~LiMn$_{1.5}$Ni$_{0.5}$O$_4$. Black thick vertical arrows inside the octahedra indicate the orientation of spin. In (a), two types of Mn ions, Mn1 and Mn2, correspond respectively to Mn$^{3+}$ and Mn$^{4+}$ that were observed experimentally. In (b), there is only one type of Mn ions which corresponds to Mn$^{4+}$, and there is one type of Ni ions which corresponds to Ni$^{2+}$. Rendered using \textsc{VESTA}~\cite{Momma:2008}.}
\label{fig:crystal_structure}
\end{figure*}

LiMn$_2$O$_4$ has been extensively studied using density-functional theory (DFT)~\cite{Hohenberg:1964, Kohn:1965} with semi-local exchange-correlation (xc) functionals augmented with Hubbard corrections, i.e. the so-called DFT+$U$ approach~\cite{anisimov:1991, Liechtenstein:1995, Dudarev:1998}. It has been shown that the on-site Hubbard $U$ correction is crucial to describe the mixed valence state of Mn ions, thus correctly predicting the existence of Mn$^{3+}$ and Mn$^{4+}$ ions and the insulating ground state of this material~\cite{Zhou:2004b, Ouyang:2009, Xu:2010, Karim:2013, Hoang:2014, Liu:2017, Isaacs:2020}. Moreover, in Ref.~\cite{Eckhoff:2020} it has been shown that various types of hybrid xc functionals correctly describe the ground-state properties of this compound. Among these, Refs.~\cite{Ouyang:2009, Liu:2017, Eckhoff:2020} investigated in detail various types of AFM ordering, eventually finding somewhat different lowest-energy spin configurations and showing that various magnetic orderings differ just by a few meV or a few tens of meV per formula unit. Conversely, first-principles studies of LiMn$_{1.5}$Ni$_{0.5}$O$_4$ are much scarcer, with Ref.~\cite{Miwa:2018} presenting a DFT+$U$ study of ferromagnetic (FM) LiMn$_{1.5}$Ni$_{0.5}$O$_4$ with a primary focus on vibrational and Raman spectral properties. In the majority of these DFT+$U$ studies of LiMn$_2$O$_4$ and LiMn$_{1.5}$Ni$_{0.5}$O$_4$ the on-site Hubbard $U$ parameter has been chosen empirically such that some experimental properties of interest are well reproduced; this is different from a fully first-principles-based approach and likely contributed to the spread of the reported results. Most importantly, the effect of inter-site Hubbard $V$ interactions~\cite{Campo:2010, TancogneDejean:2020, Lee:2020} has never been investigated in the spinel cathode materials; these interactions have been shown to be very important in other transition-metal (TM) oxides due to strong metal-ligand hybridization~\cite{Kulik:2011, Cococcioni:2019, Ricca:2020, Timrov:2020c, Mahajan:2021, Mahajan:2022, Timrov:2022c, Binci:2022, Yang:2021, Jang:2022, Yang:2022}. 

Herein, we present a fully first-principles study of the structural, electronic, magnetic, and electrochemical properties of the spinel cathode materials Li$_x$Mn$_2$O$_4$ and Li$_x$Mn$_{1.5}$Ni$_{0.5}$O$_4$ ($x=0$ and $1$) using DFT with extended Hubbard functionals (DFT+$U$+$V$)~\cite{Campo:2010}, where the on-site $U$ and inter-site $V$ Hubbard parameters are computed unbiasedly and fully from first-principles using density-functional perturbation theory (DFPT)~\cite{Timrov:2018, Timrov:2021} in a basis of L\"owdin-orthogonalized atomic orbitals. This avoids any empiricism and possible ambiguities of Hubbard-corrected DFT studies. The Hubbard parameters are determined using the self-consistent procedure prescribed in Refs.~\cite{Hsu:2009, Cococcioni:2019, Timrov:2021} to ensure that the crystal and electronic structure are mutually consistent. We find, as expected, that the on-site Hubbard $U$ correction is crucial to obtain the correct trends for various properties of these materials, but that quantitative predictions require the inclusion of the inter-site Hubbard $V$ term. This capability allows us to carefully characterize the structural and electronic properties as well as provide accurate Li-intercalation voltages in remarkable agreement with experiments. Therefore, we show that DFT+$U$+$V$ is currently the most accurate computational framework not only for the selected olivine-type cathode materials~\cite{Cococcioni:2019, Timrov:2022c} but also for the spinel cathode materials considered here, motivating further studies of other families of Li-ion cathode materials.

The paper is organized as follows: in Sec.~\ref{sec:methods} we describe the computational method; in Sec.~\ref{sec:comput_details} we summarize the computational details of the study; in Sec.~\ref{sec:results_and_discussion} we present our findings for the structural, electronic, and electrochemical properties of Li$_x$Mn$_2$O$_4$ and Li$_x$Mn$_{1.5}$Ni$_{0.5}$O$_4$ ($x=0$ and $1$) using DFT, DFT+$U$, and DFT+$U$+$V$; and in Sec.~\ref{sec:Conclusions} we present the conclusions.

\section{Computational method}
\label{sec:methods}
In this section we briefly discuss the basics of the DFT+$U$+$V$ approach~\cite{Campo:2010} and of the DFPT approach for computing Hubbard parameters~\cite{Timrov:2018, Timrov:2021}. All equations in this subsection can be easily reduced to the DFT+$U$ case by setting $V=0$. For the sake of simplicity, the formalism is presented in the framework of norm-conserving pseudopotentials in the collinear spin-polarized case. The generalization to the ultrasoft pseudopotentials  and the projector augmented wave method can be found in Ref.~\cite{Timrov:2021}. Hartree atomic units are used.
\subsection{DFT+$U$+$V$}
In DFT+$U$+$V$, a correction term is added to the approximate DFT energy functional~\cite{Campo:2010}: 
\begin{equation}
E_{\mathrm{DFT}+U+V} = E_{\mathrm{DFT}} + E_{U+V} ,
\label{eq:Edft_plus_u}
\end{equation}
where $E_{\mathrm{DFT}}$ is the approximate DFT energy (constructed, e.g., using (semi-)local exchange-correlation functionals), and $E_{U+V}$ contains the additional Hubbard term. At variance with the DFT+$U$ approach that contains only on-site interactions scaled by $U$, DFT+$U$+$V$ contains also inter-site interactions between an atom and its surrounding ligands scaled by $V$. In the case of cathode materials studied here, the on-site $U$ correction is needed for the Mn($3d$) and Ni($3d$) states, while the inter-site $V$ correction is needed for Mn($3d$)--O($2p$) and Ni($3d$)--O($2p$) interactions~\cite{Kulik:2011}. We note that the inter-site $V$ interactions for Mn($3d$)--Ni($3d$) couples are vanishing and hence these are neglected. In the simplified rotationally-invariant formulation~\cite{Dudarev:1998}, the extended Hubbard energy term reads:
\begin{eqnarray}
E_{U+V} & = & \frac{1}{2} \sum_I \sum_{\sigma m m'} 
U^I \left( \delta_{m m'} - n^{II \sigma}_{m m'} \right) n^{II \sigma}_{m' m} \nonumber \\
& & - \frac{1}{2} \sum_{I} \sum_{J (J \ne I)}^* \sum_{\sigma m m'} V^{I J} 
n^{I J \sigma}_{m m'} n^{J I \sigma}_{m' m} \,,
\label{eq:Edftu}
\end{eqnarray}
where $I$ and $J$ are atomic site indices, $m$ and $m'$ are the magnetic quantum numbers associated with a specific angular momentum [$l=2$ for Mn($3d$) and Ni($3d$), $l=1$ for O($2p$)], $U^I$ and $V^{I J}$ are the effective on-site and inter-site Hubbard parameters, and the star in the sum denotes that for each atom $I$, the index $J$ covers all its neighbors up to a given distance (or up to a given shell). 

The generalized occupation matrices $n^{I J \sigma}_{m m'}$ are computed by projecting the Kohn-Sham states on localized orbitals $\phi^{I}_{m}(\mathbf{r})$ of neighboring atoms: 
\begin{equation}
n^{I J \sigma}_{m m'} = \sum_{v,\mathbf{k}} f^\sigma_{v,\mathbf{k}}
\braket{\psi^\sigma_{v,\mathbf{k}}}{\phi^{J}_{m'}} \braket{\phi^{I}_{m}}{\psi^\sigma_{v,\mathbf{k}}} \,, 
\label{eq:occ_matrix_0}
\end{equation}
where $v$ and $\sigma$ represent, respectively, the band and spin labels of the Kohn-Sham wavefunctions $\psi^\sigma_{v,\mathbf{k}}(\mathbf{r})$, $\mathbf{k}$ indicate points in the first Brillouin zone, $f^\sigma_{v,\mathbf{k}}$ are the occupations of the Kohn-Sham states, and $\phi^I_{m}(\mathbf{r}) \equiv \phi^{\gamma(I)}_{m}(\mathbf{r} - \mathbf{R}_I)$ are localized orbitals centered on the $I$th atom of type $\gamma(I)$ at the position $\mathbf{R}_I$. It is convenient to establish a short-hand notation for the on-site occupation matrix: $n^{I\sigma}_{m m'} \equiv n^{II\sigma}_{m m'}$, which is used in the standard DFT+$U$ approach that corresponds to the first line of Eq.~\eqref{eq:Edftu}. Computing the values of $U^I$ and $V^{IJ}$ parameters is crucial to determine the degree of localization of $3d$ electrons on Mn and Ni sites and the degree of hybridization of these $3d$ electrons with $2p$ electrons centered on neighboring O sites. In the next subsection we discuss briefly how these Hubbard parameters can be computed using DFPT.

\subsection{Calculation of Hubbard parameters}
\label{sec:CalcUV_theory}
The values of Hubbard parameters are not known {\it a~priori}, and hence often these values are adjusted empirically such that the final results of simulations match some experimental properties of interest. This is fairly arbitrary, therefore, first-principles calculation of Hubbard parameters for any system at hand is essential and highly desirable. Hubbard $U$ and $V$ can be computed from a generalized piece-wise linearity condition imposed through linear-response theory~\cite{Cococcioni:2005} based on DFPT~\cite{Timrov:2018, Timrov:2021}. Within this framework the Hubbard parameters are the elements of an effective interaction matrix computed as the difference between bare and screened inverse susceptibilities~\cite{Cococcioni:2005}:
\begin{equation}
U^I = \left(\chi_0^{-1} - \chi^{-1}\right)_{II} \,,
\label{eq:Ucalc}
\end{equation}
\begin{equation}
V^{IJ} = \left(\chi_0^{-1} - \chi^{-1}\right)_{IJ} \,,
\label{eq:Vcalc}
\end{equation}
where $\chi_0$ and $\chi$ are the susceptibilities which measure the response of atomic occupations to shifts in the potential acting on individual Hubbard manifolds. In particular, $\chi$ is defined as $\chi_{IJ} = \sum_{m\sigma} \left(dn^{I \sigma}_{mm} / d\alpha^J\right)$, where $\alpha^J$ is the strength of the perturbation of electronic occupations of the $J$th site. While $\chi$ is evaluated at self-consistency of the DFPT calculation, $\chi_0$ (which has a similar definition as $\chi$) is computed before the self-consistent re-adjustment of the Hartree and exchange-correlation potentials~\cite{Timrov:2018}. In DFPT, the response of the occupation matrix is computed in a primitive unit cell as:
\begin{equation}
\frac{dn^{I \sigma}_{mm'}}{d\alpha^J} = \frac{1}{N_{\mathbf{q}}}\sum_{\mathbf{q}}^{N_{\mathbf{q}}} e^{i\mathbf{q}\cdot(\mathbf{R}_{l} - \mathbf{R}_{l'})}\Delta_{\mathbf{q}}^{s'} n^{s \sigma}_{mm'} \,,
\label{eq:dnq}
\end{equation}
where $\mathbf{q}$ is the wavevector of the monochromatic perturbation, $N_\mathbf{q}$ is the total number of $\mathbf{q}$'s, $\Delta_{\mathbf{q}}^{s'} n^{s \sigma}_{mm'}$ is the lattice-periodic response of atomic occupations to a $\mathbf{q}$-specific monochromatic perturbation, $I\equiv(l,s)$ and $J\equiv(l',s')$ where $s$ and $s'$ are the atomic indices in unit cells while $l$ and $l'$ are the unit cell indices, $\mathbf{R}_l$ and $\mathbf{R}_{l'}$ are the Bravais lattice vectors. The $\mathbf{q}$ grid is chosen fine enough to make the resulting atomic perturbations effectively decoupled from their periodic replicas. We stress that the main advantage of using DFPT is that it does not require the usage of computationally expensive supercells contrary to the original linear-response formulation of Ref.~\cite{cococcioni2005linear}. It is crucial to recall that the values of the computed Hubbard parameters strongly depend on the type of Hubbard projector functions that are used in the DFT+$U$ and DFT+$U$+$V$ approaches; this aspect is discussed in more detail in the next subsection.  

\begin{table*}[t]
\renewcommand{\arraystretch}{1.2}
\centering
\begin{tabular}{c|l|c|c|c|c|c}
\hline\hline
 \multirow{2}{*}{\parbox{1.5cm}{\centering $x$}} & \multirow{2}{*}{\parbox{1.5cm}{Method}} & \multirow{2}{*}{\parbox{1.5cm}{\centering HP}} & \multicolumn{2}{c}{Li$_x$Mn$_2$O$_4$} & \multicolumn{2}{c}{Li$_x$Mn$_{1.5}$Ni$_{0.5}$O$_4$} \\ \cline{4-7}
 &  &  & \parbox{2cm}{\centering Mn1} & \parbox{2cm}{\centering Mn2} & \parbox{2cm}{\centering Mn} & \parbox{2cm}{\centering Ni}  \\ 
\hline
  & PBEsol+$U$                      & $U$ & 6.20        & 6.52        & 6.55        & 7.53 \\ \cline{2-7}
1 & \multirow{2}{*}{PBEsol+$U$+$V$} & $U$ & 6.20        & 6.62        & 6.63        & 7.80 \\
  &                              & $V$ & $0.56-0.78$ & $0.74-0.79$ & $0.73-0.75$ & 0.65 \\ 
\hline
  & PBEsol+$U$                      & $U$ & 6.54        & 6.59        & 6.64        & 9.07 \\ \cline{2-7}
0 & \multirow{2}{*}{PBEsol+$U$+$V$} & $U$ & 6.78        & 6.80        & 6.98        & 9.39 \\
  &                              & $V$ & $0.84-0.89$ & $0.85-0.87$ & $0.90-0.94$ & 0.86 \\
\hline\hline
\end{tabular}
\caption{Self-consistent Hubbard parameters (HP) in eV for Li$_x$Mn$_2$O$_4$ and Li$_x$Mn$_{1.5}$Ni$_{0.5}$O$_4$ at $x=1$ and $x=0$ computed using DFPT (PBEsol functional) in the basis of L\"owdin-orthogonalized atomic orbitals as Hubbard projector functions.}
\label{tab:Hub_param}
\end{table*}

\subsection{Hubbard projectors}
\label{sec:Hub_projectors_theory}
The Hubbard manifold $\{ \phi_m^I(\mathbf{r}) \}$ can be constructed using different types of projector functions (see e.g. Refs.~\cite{Tablero:2008, Timrov:2020b}). Here we consider atomic orbitals that are orthogonalized using the L\"owdin method~\cite{Lowdin:1950, Mayer:2002}:
\begin{equation}
    \phi^I_{m}(\mathbf{r}) = \sum_{J m'} \left(\hat{O}^{-\frac{1}{2}}\right)^{JI}_{m' m} \varphi^J_{m'}(\mathbf{r}) \,,
    \label{eq:OAO_def}
\end{equation}
where $\varphi^I_{m}(\mathbf{r})$ are the nonorthogonalized atomic orbitals, $\hat{O}$ is the orbital overlap matrix which is defined as $(\hat{O})^{IJ}_{m_1 m_2} = \braket{\varphi^I_{m_1}}{\varphi^J_{m_2}}$, and $(\hat{O})^{IJ}_{m_1 m_2}$ is a matrix element of $\hat{O}$. This basis set better represents hybridizations of orbitals between neighboring sites, but especially it allows us to avoid counting Hubbard corrections twice in the interstitial regions between atoms. It is important to note that L\"owdin-orthogonalized atomic orbitals are not truncated at some cutoff radius (at variance with other implementations, Refs.~\cite{Amadon:2008, Nawa:2018}), which thus eliminates ambiguities due to the choice of such a cutoff radius~\cite{Timrov:2020b, Wang:2016}.

\section{Computational details}
\label{sec:comput_details}

All calculations are performed using the \textsc{Quantum ESPRESSO} distribution~\cite{Giannozzi:2009, Giannozzi:2017, Giannozzi:2020}. For LiMn$_2$O$_4$, a supercell composed of 8 formula units (56 atoms) is used in order to model an AFM collinear magnetic ordering with a space group $Imma$, following Ref.~\cite{Liu:2017}. This supercell is a pseudo-cubic cell that is built starting from an orthorhombic spinel structure~\cite{Kuwabara:2012}. The AFM ordering is chosen to be the lowest-energy one found in Ref.~\cite{Liu:2017}, and is shown in Fig.~\ref{fig:crystal_structure}(a). Two types of Mn ions, labeled here as Mn1 and Mn2, correspond respectively to Mn$^{3+}$ and Mn$^{4+}$ that were observed experimentally. For LiMn$_{1.5}$Ni$_{0.5}$O$_4$, a supercell composed of 8 formula units (56 atoms) is also used in order to model a FiM collinear magnetic ordering~\cite{Amdouni:2007, MoorheadRosenberg:2012, MoorheadRosenberg:2013} with a space group $Fd\bar{3}m$, and is shown in Fig.~\ref{fig:crystal_structure}(b). We have checked that different cation orderings within $Fd\bar{3}m$ lead to the same electronic structure. The delithiated materials, Mn$_2$O$_4$ and Mn$_{1.5}$Ni$_{0.5}$O$_4$, are modelled using the same supercells and magnetic orderings as described above, and are constructed by simply removing all Li atoms. It is worth mentioning that Mn$_2$O$_4$ corresponds to the $\lambda$ phase of MnO$_2$~\cite{Wills:1999, Kitchaev:2016}. 

For the xc functional we use a PBEsol~\cite{Perdew:2008} spin-polarized generalized-gradient approximation~\cite{Note_PBEsol}. Pseudopotentials are those of the SSSP library~v1.1 (precision)~\cite{prandini2018precision, MaterialsCloud}: \texttt{mn\_pbesol\_v1.5.uspp.F.UPF} (GBRV library v1.5~\cite{Garrity:2014}), \texttt{O.pbesol-n-kjpaw\_psl.0.1.UPF} (Pslibrary v0.3.1~\cite{Kucukbenli:2014}), \texttt{ni\_pbesol\_v1.4.uspp.F.UPF} and \texttt{li\_pbesol\_v1.4.uspp.F.UPF} (GBRV library v1.4~\cite{Garrity:2014}). For metallic ground states, we use the Marzari-Vanderbilt (MV) smearing~\cite{Marzari:1999} with a broadening parameter of $0.01$~Ry. The crystal structure for all spin configurations is optimized at three levels of theory (PBEsol, PBEsol+$U$, and PBEsol+$U$+$V$) using the Broyden-Fletcher-Goldfarb-Shanno (BFGS) algorithm~\cite{Fletcher:1987} with convergence thresholds of $10^{-6}$~Ry, $10^{-5}$~Ry/Bohr, and $0.5$~KBar for the total energy, forces, and pressure, respectively. For structural optimizations, the $\mathbf{k}$-point sampling of the Brillouin zone uses a uniform $\Gamma$-centered $6 \times 6 \times 6$ mesh. Kohn-Sham wavefunctions and charge density are expanded in plane waves up to a kinetic-energy cutoff of 90 and 1080~Ry, respectively. Projected density of states (PDOS) is plotted using a Gaussian smearing with a broadening parameter of $10^{-3}$~Ry.

In order to compute the Li-intercalation voltages in cathode materials, it is necessary to compute the total energy of bulk Li; this is modelled using PBEsol and the \textit{bcc} unit cell with one Li atom at the origin. The optimized lattice parameter is 3.436~\AA, the Brillouin zone is sampled using the uniform $\Gamma$-centered $10 \times 10 \times 10$ $\mathbf{k}$-point mesh, and we use the MV smearing with a broadening of 0.02~Ry. The Kohn-Sham wavefunctions and charge density are expanded in plane waves up to a kinetic-energy cutoff of 65 and 780~Ry, respectively.

PBEsol+$U$ and PBEsol+$U$+$V$ calculations are performed using the L\"owdin-orthogonalized atomic orbitals as Hubbard projectors~\cite{Lowdin:1950, Mayer:2002, Timrov:2020b}. Hubbard $U$ and $V$ parameters are computed using DFPT~\cite{Timrov:2018, Timrov:2021} as implemented in the \textsc{HP} code~\cite{Timrov:2022} which is part of \textsc{Quantum ESPRESSO}. It is worth noting that computationally expensive summations over empty states in perturbation theory are avoided thanks to the use of projectors on empty states manifolds (see e.g. Refs.~\cite{Baroni:2001, Gorni:2018}). We have used uniform $\Gamma$-centered $\mathbf{k}$- and $\mathbf{q}$-point meshes of size $4 \times 4 \times 4$ and $2 \times 2 \times 2$, respectively. Kohn-Sham wavefunctions and charge density are expanded in plane waves up to a kinetic-energy cutoff of 65 and 780~Ry, respectively, with an accuracy in the computed Hubbard parameters of $\sim 0.01$~eV. It is important to stress that we have used a self-consistent procedure for the calculation of $U$ and $V$, as described in detail in Ref.~\cite{Timrov:2021}, which consists of cyclic calculations containing structural optimizations and recalculations of Hubbard parameters for each new geometry. This procedure has proven to provide accurate results for various TM compounds~\cite{Ricca:2019, Floris:2020, Sun:2020, Zhou:2021, KirchnerHall:2021, Xiong:2021}. The resulting Hubbard parameters are presented in Table~\ref{tab:Hub_param}. It can be seen that the values of Hubbard $U$ for Mn and Ni ions are sensitive to their oxidation state (OS) (see Sec.~\ref{sec:Oxidation_state}), in line with previous studies for phospho-olivines~\cite{Cococcioni:2019, Timrov:2022c}. Hubbard $V$ values show changes of only $\sim 0.1-0.2$~eV upon (de-)lithiation due to changes in the metal-ligand interatomic distances and electronic screening.

The data used to produce the results of this paper are available in the Materials Cloud Archive~\cite{MaterialsCloudArchive2023}.

\section{Results and discussion}
\label{sec:results_and_discussion}

\subsection{Structural properties}

Tables~\ref{tab:lattice_param_LMO} and \ref{tab:lattice_param_LMNO} compare the lattice parameters and cell volumes for Li$_x$Mn$_2$O$_4$ and Li$_x$Mn$_{1.5}$Ni$_{0.5}$O$_4$ ($x=0$ and 1) optimized using PBEsol, PBEsol+$U$, and PBEsol+$U$+$V$ as well as the experimental values~\cite{Kanno:1999, Huang:2011, Jang:2000}. For LiMn$_2$O$_4$, PBEsol underestimates the cell volume by 7.7\%, while PBEsol+$U$ overestimates it by 2.4\%, and PBEsol+$U$+$V$ provides the most accurate prediction with a small overestimation of 1.2\%. For Mn$_2$O$_4$, the trend is different. Namely, PBEsol predicts the cell volume which is in closest agreement with experimental values and is underestimated by 2.7\%, while PBEsol+$U$ overestimates the cell volume by 6.6\%, and PBEsol+$U$+$V$ also overestimates it but by a smaller margin of 3.9\%. 

In the case of partial substitution of Ni for Mn, the trends are the same but more pronounced. For LiMn$_{1.5}$Ni$_{0.5}$O$_4$, PBEsol underestimates the cell volume by 3.2\%, while PBEsol+$U$ overestimates it by 4.1\%, and PBEsol+$U$+$V$ again provides the most accurate prediction with an overestimation of 2.7\% (which is about as twice large as the case of LiMn$_2$O$_4$). For Mn$_{1.5}$Ni$_{0.5}$O$_4$, again PBEsol provides remarkable agreement with the experimental cell volume with an underestimation of only 1.0\%, while both PBEsol+$U$ and PBEsol+$U$+$V$ show quite a large overestimation by 6.5\% and 6.2\%, respectively. Importantly, only using PBEsol+$U$ and PBEsol+$U$+$V$ we find that the cell volume is decreased by partially substituting Ni for Mn in LiMn$_2$O$_4$ in agreement with experiments, while PBEsol incorrectly shows the opposite trend. In the fully delithiated material, conversely, all levels of theory are consistent with experiments and show that the cell volume decreases upon the partial substitution of Ni for Mn. 

In summary, while PBEsol+$U$+$V$ predicts the crystal structure of fully lithiated spinel cathodes most accurately, it is still less accurate than PBEsol for the fully delithiated structures, in line with similar findings for phospho-olivines~\cite{Cococcioni:2019, Timrov:2022c}. Relatively large overestimations of the cell volume in the delithiated structures are mainly driven by the large on-site Hubbard $U$ correction, while inter-site $V$ corrections counteract it and reduce this effect. The effect of inter-site Hubbard $V$ interactions, therefore, is an essential correction on top of PBEsol+$U$ to provide an improved description of the structural properties of the spinel cathode materials.

\begin{table}[t]
\renewcommand{\arraystretch}{1.2}
\centering
\begin{tabular}{lccc}
\hline\hline
Method & \parbox{1cm}{\centering LP}  & \parbox{2.2cm}{\centering LiMn$_2$O$_4$} & \parbox{2.2cm}{\centering Mn$_2$O$_4$} \\ 
\hline
\multirow{4}{*}{PBEsol}         & $a$ (\AA)     &   8.06       &  7.99      \\
                             & $b$ (\AA)     &   8.06       &  7.99      \\
                             & $c$ (\AA)     &   8.08       &  7.98      \\
                             & $V$ (\AA$^3$) &   525.0      &  509.6     \\ 
\hline
\multirow{4}{*}{PBEsol+$U$}     & $a$ (\AA)     &   8.18       &  8.24      \\
                             & $b$ (\AA)     &   8.18       &  8.24      \\
                             & $c$ (\AA)     &   8.72       &  8.23      \\
                             & $V$ (\AA$^3$) &   582.7      &  558.2     \\
\hline
\multirow{4}{*}{PBEsol+$U$+$V$} & $a$ (\AA)     &   8.13       &  8.17      \\
                             & $b$ (\AA)     &   8.13       &  8.17      \\
                             & $c$ (\AA)     &   8.71       &  8.16      \\
                             & $V$ (\AA$^3$) &   575.5      &  544.2     \\
\hline
\multirow{4}{*}{Expt.~\cite{Kanno:1999, Huang:2011, Jang:2000}}      
                             & $a$ (\AA)     &   8.11       &  8.06      \\
                             & $b$ (\AA)     &   8.11       &  8.06      \\
                             & $c$ (\AA)     &   8.65       &  8.06      \\
                             & $V$ (\AA$^3$) &   568.9      &  523.6     \\
\hline\hline
\end{tabular}
\caption{Pseudo-cubic lattice parameters ($a$, $b$, and $c$) and cell volume ($V$) of Li$_x$Mn$_2$O$_4$ at $x=1$ and $x=0$.}
\label{tab:lattice_param_LMO}
\end{table}

\begin{table}[t]
\renewcommand{\arraystretch}{1.2}
\centering
\begin{tabular}{lccc}
\hline\hline
Method & \parbox{1cm}{\centering LP}  & \parbox{2.2cm}{\centering LiMn$_{1.5}$Ni$_{0.5}$O$_4$} & \parbox{2.2cm}{\centering Mn$_{1.5}$Ni$_{0.5}$O$_4$} \\ 
\hline
\multirow{2}{*}{PBEsol}         & $a$ (\AA)     &  8.08    &  7.97       \\
                             & $V$ (\AA$^3$) &  528.1   &  506.8      \\ 
\hline
\multirow{2}{*}{PBEsol+$U$}     & $a$ (\AA)     &  8.28    &  8.17       \\
                             & $V$ (\AA$^3$) &  567.8   &  545.2      \\
\hline
\multirow{2}{*}{PBEsol+$U$+$V$} & $a$ (\AA)     &  8.24    &  8.16       \\
                             & $V$ (\AA$^3$) &  560.0   &  543.6      \\
                             \hline
\multirow{2}{*}{Expt.~\cite{Kan:2017, Amin:2020}}       
                             & $a$ (\AA)     &  8.17    &  8.00       \\
                             & $V$ (\AA$^3$) &  545.3   &  512.0      \\
\hline\hline
\end{tabular}
\caption{Cubic lattice parameter ($a$) and cell volume ($V$) of Li$_x$Mn$_{1.5}$Ni$_{0.5}$O$_4$ at $x=1$ and $x=0$.}
\label{tab:lattice_param_LMNO}
\end{table}

\subsection{L\"owdin occupations, magnetic moments, and oxidation state}
\label{sec:Oxidation_state}

As mentioned in the introduction, in LiMn$_2$O$_4$, Mn exists in two OS, namely Mn$^{3+}$ and Mn$^{4+}$, in an equal $1:1$ ratio. Upon delithiation, each Mn$^{3+}$ ion loses one electron and thus oxidizes to Mn$^{4+}$, and as a result in Mn$_2$O$_4$ all Mn ions are in the $+4$ OS~\cite{Wills:1999}. On the other hand, in LiMn$_{1.5}$Ni$_{0.5}$O$_4$ all Mn ions are in the $+4$ OS while all Ni ions are in the $+2$ OS (the ratio of Ni to Mn is $1:3$). Upon delithiation, each Ni ion loses two electrons and as a consequence they all oxidize from Ni$^{2+}$ to Ni$^{4+}$, while the Mn$^{4+}$ ions remain unchanged~\cite{Manthiram:2014}.  

\begin{table*}[t]
\centering
\begin{tabular}{c|l|c|ccccccccccccc}
\hline\hline
Material & Method & Type & $\lambda_1^{\uparrow}$ & $\lambda_2^{\uparrow}$ & $\lambda_3^{\uparrow}$ & $\lambda_4^{\uparrow}$ & $\lambda_5^{\uparrow}$ &  $\lambda_1^{\downarrow}$ & $\lambda_2^{\downarrow}$ & $\lambda_3^{\downarrow}$ & $\lambda_4^{\downarrow}$ & $\lambda_5^{\downarrow}$ & $n$ & $m$ ($\mu_\mathrm{B}$) & OS \\
\hline
\parbox[t]{7mm}{\multirow{8}{*}{\rotatebox[origin=c]{90}{LiMn$_2$O$_4$}}} 
& \multirow{2}{*}{PBEsol}         & Mn1 & 0.47  &      0.49  & {\bf 0.99} & {\bf 0.99} & {\bf 0.99} & 0.18 & 0.24 & 0.27 & 0.33 & 0.33  & 5.25  &  2.58 & +4 \\
&                              & Mn2 & 0.44  &      0.44  & {\bf 0.98} & {\bf 0.99} & {\bf 0.99} & 0.15 & 0.26 & 0.31 & 0.35 & 0.38  & 5.28  &  2.39 & +4 \\ \cline{2-16}
& \multirow{2}{*}{PBEsol+$U$}     & Mn1 & 0.56  & {\bf 0.99} & {\bf 1.00} & {\bf 1.00} & {\bf 1.00} & 0.04 & 0.04 & 0.06 & 0.14 & 0.22  & 5.02  &  4.05 & +3 \\
&                              & Mn2 & 0.64  &      0.65  & {\bf 0.99} & {\bf 1.00} & {\bf 1.00} & 0.06 & 0.07 & 0.07 & 0.24 & 0.27  & 4.98  &  3.57 & +4 \\ \cline{2-16}
& \multirow{2}{*}{PBEsol+$U$+$V$} & Mn1 & 0.52  & {\bf 0.99} & {\bf 1.00} & {\bf 1.00} & {\bf 1.00} & 0.04 & 0.05 & 0.07 & 0.15 & 0.24  & 5.05  &  3.95 & +3 \\
&                              & Mn2 & 0.59  &      0.60  & {\bf 1.00} & {\bf 1.00} & {\bf 1.00} & 0.08 & 0.08 & 0.09 & 0.27 & 0.30  & 5.00  &  3.37 & +4 \\ \cline{2-16}  
& \multirow{2}{*}{Nominal}     & Mn1 & 0.00  & {\bf 1.00} & {\bf 1.00} & {\bf 1.00} & {\bf 1.00} & 0.00 & 0.00 & 0.00 & 0.00 & 0.00  & 4.00  &  4.00 & +3 \\
&                              & Mn2 & 0.00  &      0.00  & {\bf 1.00} & {\bf 1.00} & {\bf 1.00} & 0.00 & 0.00 & 0.00 & 0.00 & 0.00  & 3.00  &  3.00 & +4 \\
\hline
\parbox[t]{7mm}{\multirow{8}{*}{\rotatebox[origin=c]{90}{Mn$_2$O$_4$}}} 
& \multirow{2}{*}{PBEsol}         & Mn1 & 0.49  &      0.50  & {\bf 0.98} & {\bf 0.98} & {\bf 0.99} & 0.15 & 0.16 & 0.20 & 0.36 & 0.37  & 5.16  &  2.71 & +4 \\
&                              & Mn2 & 0.49  &      0.51  & {\bf 0.98} & {\bf 0.98} & {\bf 0.99} & 0.15 & 0.16 & 0.19 & 0.34 & 0.37  & 5.16  &  2.75 & +4 \\ \cline{2-16}
& \multirow{2}{*}{PBEsol+$U$}     & Mn1 & 0.64  &      0.65  & {\bf 1.00} & {\bf 1.00} & {\bf 1.00} & 0.06 & 0.06 & 0.08 & 0.26 & 0.26  & 4.99  &  3.55 & +4 \\
&                              & Mn2 & 0.65  &      0.67  & {\bf 0.99} & {\bf 0.99} & {\bf 1.00} & 0.06 & 0.06 & 0.08 & 0.24 & 0.25  & 4.99  &  3.63 & +4 \\ \cline{2-16}
& \multirow{2}{*}{PBEsol+$U$+$V$} & Mn1 & 0.59  &      0.60  & {\bf 1.00} & {\bf 1.00} & {\bf 1.00} & 0.07 & 0.07 & 0.10 & 0.30 & 0.30  & 5.02  &  3.33 & +4 \\
&                              & Mn2 & 0.60  &      0.62  & {\bf 0.99} & {\bf 1.00} & {\bf 1.00} & 0.07 & 0.07 & 0.10 & 0.28 & 0.29  & 5.01  &  3.40 & +4 \\ \cline{2-16}  
& \multirow{2}{*}{Nominal}     & Mn1 & 0.00  &      0.00  & {\bf 1.00} & {\bf 1.00} & {\bf 1.00} & 0.00 & 0.00 & 0.00 & 0.00 & 0.00  & 3.00  &  3.00 & +4 \\
&                              & Mn2 & 0.00  &      0.00  & {\bf 1.00} & {\bf 1.00} & {\bf 1.00} & 0.00 & 0.00 & 0.00 & 0.00 & 0.00  & 3.00  &  3.00 & +4 \\
\hline
\parbox[t]{7mm}{\multirow{8}{*}{\rotatebox[origin=c]{90}{LiMn$_{1.5}$Ni$_{0.5}$O$_4$}}} 
& \multirow{2}{*}{PBEsol}         & Mn & 0.48  &      0.48  & {\bf 0.99} & {\bf 0.99} & {\bf 0.99} &      0.15  &      0.17  &      0.26  &      0.34  &      0.34   & 5.18  &   2.68 & +4  \\
&                              & Ni & 0.35  &      0.35  & {\bf 0.97} & {\bf 0.97} & {\bf 0.97} &      0.87  &      0.87  & {\bf 0.99} & {\bf 0.99} & {\bf 0.99}  & 8.29  &  -1.08 & +4  \\ \cline{2-16}
& \multirow{2}{*}{PBEsol+$U$}     & Mn & 0.64  &      0.66  & {\bf 1.00} & {\bf 1.00} & {\bf 1.00} &      0.06  &      0.06  &      0.08  &      0.23  &      0.26   & 4.98  &   3.61 & +4  \\
&                              & Ni & 0.14  &      0.14  & {\bf 0.98} & {\bf 0.99} & {\bf 0.99} & {\bf 0.99} & {\bf 0.99} & {\bf 1.00} & {\bf 1.00} & {\bf 1.00}  & 8.21  &  -1.74 & +2 \\ \cline{2-16}
& \multirow{2}{*}{PBEsol+$U$+$V$} & Mn & 0.60  &      0.62  & {\bf 1.00} & {\bf 1.00} & {\bf 1.00} &      0.07  &      0.07  &      0.10  &      0.27  &      0.29   & 5.00  &   3.41 & +4 \\
&                              & Ni & 0.15  &      0.15  & {\bf 0.99} & {\bf 0.99} & {\bf 0.99} & {\bf 0.99} & {\bf 0.99} & {\bf 0.99} & {\bf 1.00} & {\bf 1.00}  & 8.24  &  -1.71 & +2 \\ \cline{2-16}  
& \multirow{2}{*}{Nominal}     & Mn & 0.00  &      0.00  & {\bf 1.00} & {\bf 1.00} & {\bf 1.00} &      0.00  &      0.00  &      0.00  &      0.00  &      0.00   & 3.00  &   3.00 & +4 \\
&                              & Ni & 0.00  &      0.00  & {\bf 1.00} & {\bf 1.00} & {\bf 1.00} & {\bf 1.00} & {\bf 1.00} & {\bf 1.00} & {\bf 1.00} & {\bf 1.00}  & 8.00  &  -2.00 & +2 \\
\hline
\parbox[t]{7mm}{\multirow{8}{*}{\rotatebox[origin=c]{90}{Mn$_{1.5}$Ni$_{0.5}$O$_4$}}} 
& \multirow{2}{*}{PBEsol}         & Mn & 0.51  &      0.51  & {\bf 0.97} & {\bf 0.99} & {\bf 0.99} &      0.13  &      0.15  &      0.19  &      0.35  &      0.36   & 5.14  &   2.79 & +4  \\
&                              & Ni & 0.59  &      0.59  & {\bf 0.99} & {\bf 0.99} & {\bf 0.99} &      0.52  &      0.52  & {\bf 0.98} & {\bf 0.98} & {\bf 0.99}  & 8.15  &   0.16 & +4  \\ \cline{2-16}
& \multirow{2}{*}{PBEsol+$U$}     & Mn & 0.66  &      0.67  & {\bf 0.99} & {\bf 1.00} & {\bf 1.00} &      0.05  &      0.06  &      0.08  &      0.24  &      0.25   & 4.99  &   3.64 & +4  \\
&                              & Ni & 0.42  &      0.42  & {\bf 0.99} & {\bf 0.99} & {\bf 0.99} &      0.68  &      0.68  & {\bf 1.00} & {\bf 1.00} & {\bf 1.00}  & 8.16  &  -0.53 & +4 \\ \cline{2-16}
& \multirow{2}{*}{PBEsol+$U$+$V$} & Mn & 0.62  &      0.63  & {\bf 0.99} & {\bf 1.00} & {\bf 1.00} &      0.06  &      0.07  &      0.09  &      0.27  &      0.28   & 5.00  &   3.47 & +4 \\
&                              & Ni & 0.81  &      0.81  & {\bf 1.00} & {\bf 1.00} & {\bf 1.00} &      0.28  &      0.28  & {\bf 1.00} & {\bf 1.00} & {\bf 1.00}  & 8.13  &   1.06 & +4 \\ \cline{2-16}  
& \multirow{2}{*}{Nominal}     & Mn & 0.00  &      0.00  & {\bf 1.00} & {\bf 1.00} & {\bf 1.00} &      0.00  &      0.00  &      0.00  &      0.00  &      0.00   & 3.00  &   3.00 & +4 \\
&                              & Ni & 0.00  &      0.00  & {\bf 1.00} & {\bf 1.00} & {\bf 1.00} &      0.00  &      0.00  & {\bf 1.00} & {\bf 1.00} & {\bf 1.00}  & 6.00  &   0.00 & +4 \\
\hline\hline
\end{tabular}
\caption{L\"owdin population analysis data for the $3d$ shell of Mn and Ni ions in Li$_x$Mn$_2$O$_4$ and LiMn$_{1.5}$Ni$_{0.5}$O$_4$ at $x=0$ and $x=1$ computed using three approaches: PBEsol, PBEsol+$U$, and PBEsol+$U$+$V$. We compare these values with the expectations from using the nominal valence of the TM ions. This table shows the eigenvalues of the site-diagonal occupation matrix for the spin-up ($\lambda_i^\uparrow$, $i=\overline{1,5}$) and spin-down ($\lambda_i^\downarrow$, $i=\overline{1,5}$) channels, L\"owdin occupations $n = \sum_i (\lambda_i^\uparrow + \lambda_i^\downarrow)$, magnetic moments $m = \sum_i (\lambda_i^\uparrow - \lambda_i^\downarrow)$, and the oxidation state (OS). In LiMn$_2$O$_4$ and Mn$_2$O$_4$, which are both AFM, the magnetic moments of Mn1 and Mn2 have both positive and negative values (see Fig.~\ref{fig:crystal_structure}(a)), and here we report only the positive values. The eigenvalues are written in the ascending order (from left to right) for each spin channel. The eigenvalues written in bold correspond to fully occupied states and thus are taken into account when determining the OS according to Ref.~\cite{Sit:2011}.}
\label{tab:OS}
\end{table*}

\begin{figure*}[t]
  \centering
  \includegraphics[width=0.85\linewidth]{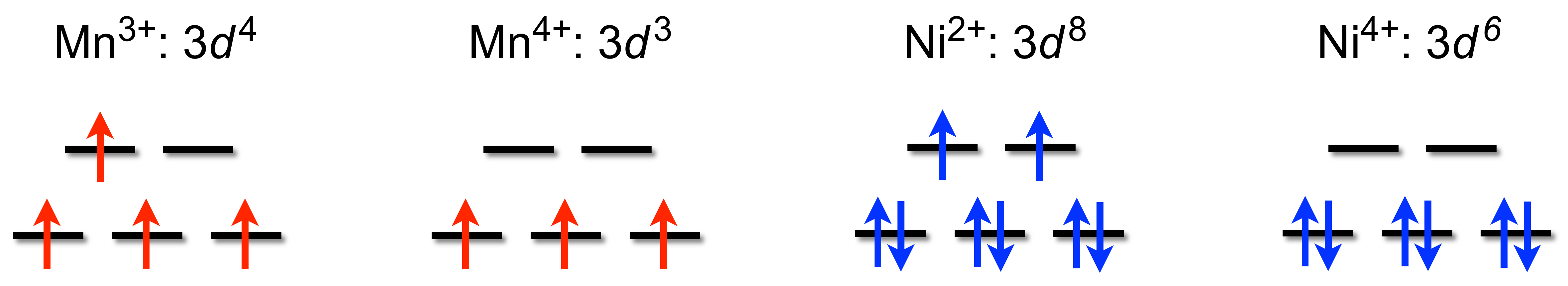}
   \caption{Nominal occupations of the $3d$ manifold of Mn and Ni atoms (not hybridized with ligands) in an undistorted octahedral complex with different oxidation states ($O_h$ point group). The $t_{2g}$ (three lower) and $e_g$ (two higher) levels are indicated with black horizontal lines and are nondegenerate due to the crystal-field splitting; up and down arrows correspond to spin-up and spin-down electrons, respectively.}
\label{fig:Nominal_OS} 
\end{figure*}

Theoretical determination of the OS of TM ions based on (Hubbard-corrected) PBEsol calculations is not a trivial task, and can be carried out using different methods (see e.g. Ref.~\cite{Timrov:2022c} for a brief review). In particular, the atomic occupations---that are computed by projecting the Kohn-Sham wavefunctions on a certain type of atom-centered localized orbitals (and thus are not uniquely defined)---are often used to determine the OS, but this approach can lead to the wrong conclusion due to the so-called negative-feedback charge regulation mechanism~\cite{Raebiger:2008, Resta:2008}. Another popular proxy that is commonly used to determine the OS of TM ions are magnetic moments. However, magnetic moments are also not uniquely defined: they are either computed by integrating the magnetization density inside the spheres centered on atoms (and hence depend on the cutoff radius~\cite{Reed:2002}) or computed via difference of the spin-up and spin-down components of the Kohn-Sham wavefunctions projected on atom-centered localized functions (and hence depend on the specific type of these localized functions). Additionally, a given OS can have multiple spin configurations (e.g., high or low spin) which have distinct magnetic moments.
Since none of these methods are very reliable~\cite{Timrov:2022c}, we chose to use the projection-based method proposed by Sit and coworkers that uses eigenvalues of the atomic occupation matrix to determine the OS of TM ions~\cite{Sit:2011}.

Table~\ref{tab:OS} presents the spin-resolved eigenvalues of the atomic occupations matrices, L\"owdin occupations, magnetic moments, and OSs computed using PBEsol, PBEsol+$U$, and PBEsol+$U$+$V$ for all materials studied here. According to Ref.~\cite{Sit:2011}, in order to determine the OS, we only need to count the eigenvalues that are close to 1.0, and all others (distinctly lower than 1) must be disregarded since they describe electrons that are shared between TM ions and ligands -- here, the O-$2p$ states -- and hence do not belong exclusively to the former. Moreover, for the sake of clarity, in Fig.~\ref{fig:Nominal_OS} we show how many electrons must be in the $3d$ shells of TM ions in order to correspond to Mn$^{3+}$, Mn$^{4+}$, Ni$^{2+}$, and Ni$^{4+}$. We can see from Table~\ref{tab:OS}, that only PBEsol+$U$ and PBEsol+$U$+$V$ correctly predict the OS of Mn and Ni ions in the fully lithiated and delithiated spinel materials, while PBEsol fails and incorrectly  predicts the $+4$ OS for Mn1 and Ni in the lithiated compounds. This failure of PBEsol is due to the overdelocalization of Mn-$3d$ and Ni-$3d$ states originating from self-interaction errors inherent to local and semi-local xc functionals~\cite{Perdew:1981, MoriSanchez:2006}. These errors are successfully alleviated thanks to Hubbard $U$ and $V$ corrections~\cite{Kulik:2006, Kulik:2008, Kulik:2011}. Therefore, both within PBEsol+$U$ and PBEsol+$U$+$V$ we find Mn$^{4+}$ and high-spin Mn$^{3+}$ in LiMn$_2$O$_4$, and only Mn$^{4+}$ in Mn$_2$O$_4$; and in LiMn$_{1.5}$Ni$_{0.5}$O$_4$ we find Mn$^{4+}$ and Ni$^{2+}$, while in Mn$_{1.5}$Ni$_{0.5}$O$_4$ we find Mn$^{4+}$ and low-spin Ni$^{4+}$.

It is useful here to comment on L\"owdin occupations. Indeed, due to the negative-feedback charge regulation mechanism~\cite{Raebiger:2008, Resta:2008}, in Li$_x$Mn$_2$O$_4$ we find too small changes in the L\"owdin occupations upon delithiation (from $x=1$ to $x=0$); more specifically, within PBEsol+$U$+$V$ the occupations for Mn1 ions change only from 5.05 to 5.02, while the nominal occupations change from 4.00 (Mn$^{3+}$) to 3.00 (Mn$^{4+}$). Even though the absolute values of L\"owdin occupations are not very reliable due to their strong dependence on the type of atom-centered localized functions on which one projects the Kohn-Sham wavefunctions, we highlight here that the lack of the correct change in the L\"owdin occupations upon delithiation makes it impossible to determine the OS, as previously mentioned. Similar observations can be made for the L\"owdin occupations of Ni ions in Li$_x$Mn$_{1.5}$Ni$_{0.5}$O$_4$ upon delithiation. Namely, within PBEsol+$U$+$V$ the occupations for Ni ions change only from 8.24 to 8.13, while the nominal occupations change from 8.00 (Ni$^{2+}$) to 6.00 (Ni$^{4+}$). Similar trends are observed also within PBEsol and PBEsol+$U$ confirming that L\"owdin occupations are not a good proxy for determining the OS.

Lastly, let us analyze the magnetic moments. Within PBEsol+$U$+$V$ for Li$_x$Mn$_2$O$_4$, the magnetic moments for Mn1 ions change from 3.95 to 3.33~$\mu_\mathrm{B}$ upon delithiation, while the nominal magnetic moments change from 4.00 (Mn$^{3+}$) to 3.00~$\mu_\mathrm{B}$ (Mn$^{4+}$). Therefore, even though the absolute values of magnetic moments for Mn ions do not match precisely the nominal ones (and this is correct due to orbital hybridization effects in the crystal compared to an isolated atom in an idealized undistorted octahedral complex, see Fig.~\ref{fig:Nominal_OS}), it is quite straightforward to make correct attributions of OSs to Mn1 ions to be Mn$^{3+}$ and Mn$^{4+}$ in LiMn$_2$O$_4$ and Mn$_2$O$_4$, respectively. However, such an analysis is much less clear in the case of Li$_x$Mn$_{1.5}$Ni$_{0.5}$O$_4$. Indeed, within PBEsol+$U$+$V$ the magnetic moments of Ni ions change from $-1.71$ to $1.06$~$\mu_\mathrm{B}$ upon delithiation, while the nominal magnetic moments change from $-2.00$ (Ni$^{2+}$) to $0.00$~$\mu_\mathrm{B}$ (Ni$^{4+}$). Neither PBEsol nor PBEsol+$U$ are able to show changes in magnetic moments for Ni ions similar to the nominal ones. This is due to the fact that the partial occupancy of the ``formally empty'' (but hybridized in practice) spin-up and spin-down channels of Ni ions are very sensitive to the Hubbard corrections~\cite{Binci:2022}. Therefore, it turns out that the magnetic moments are not always a reliable proxy for determining the OS of TM ions. Finally, for the sake of completeness, we remark that at all levels of theory the total net magnetization is exactly zero for LiMn$_2$O$_4$ and Mn$_2$O$_4$, while it is 3.5 and 4.5~$\mu_\mathrm{B}$ per formula unit for LiMn$_{1.5}$Ni$_{0.5}$O$_4$ and Mn$_{1.5}$Ni$_{0.5}$O$_4$, respectively, in agreement with theoretical nominal net magnetizations~\cite{MoorheadRosenberg:2012, MoorheadRosenberg:2013}.

\subsection{Band gaps}
\label{sec:Band_gaps}

From experimental observations it is known that LiMn$_2$O$_4$ is insulating with a band gap of $\sim 1.2$~eV~\cite{Ouyang:2006}; however, we are not aware of any experimental reports for Mn$_2$O$_4$ ($\lambda$-MnO$_2$). Table~\ref{tab:band_gaps_LMO} summarizes the computed band gaps for these two compounds using PBEsol, PBEsol+$U$, and PBEsol+$U$+$V$. It can be seen for LiMn$_2$O$_4$ that PBEsol predicts a metallic ground state, PBEsol+$U$ underestimated the gap by 39\%, while PBEsol+$U$+$V$ predicts a band gap of 1.19~eV in excellent agreement with the experimental value of Ref.~\cite{Ouyang:2006} and in good agreement with the value of 1.1~eV computed using the PBE0r hybrid functional~\cite{Eckhoff:2020}. This agrees with the previous findings that PBEsol+$U$ often improves the band gaps compared to PBEsol~\cite{KirchnerHall:2021, Xiong:2021}, while PBEsol+$U$+$V$ outperforms PBEsol+$U$ in terms of accuracy for band gaps~\cite{Mahajan:2021, Mahajan:2022}. For Mn$_2$O$_4$, PBEsol predicts the smallest band gap, PBEsol+$U$+$V$ the largest, while PBEsol+$U$ provides the band gap in between. It is worth noting that the PBEsol+$U$+$V$ value of 2.05~eV is close to the band gap value of 2.2~eV that was computed using PBE0r~\cite{Eckhoff:2020}, thus suggesting that these values should be reliable. 

Unfortunately, for LiMn$_{1.5}$Ni$_{0.5}$O$_4$ and Mn$_{1.5}$Ni$_{0.5}$O$_4$ there are no experimental data for the band gaps, to the best of our knowledge; however, the work of Ref.~\cite{Singh:2019} states that LiMn$_{1.5}$Ni$_{0.5}$O$_4$ has an insulating ground state. Table~\ref{tab:band_gaps_LMNO} presents the band gaps for these two compounds computed using PBEsol, PBEsol+$U$, and PBEsol+$U$+$V$. Like in the case of LiMn$_2$O$_4$, for LiMn$_{1.5}$Ni$_{0.5}$O$_4$ PBEsol predicts a metallic ground state, while the fully delithiated counterpart (i.e. Mn$_{1.5}$Ni$_{0.5}$O$_4$) has some finite but very small band gap of 0.37~eV. On the other hand, PBEsol+$U$ opens a gap in LiMn$_{1.5}$Ni$_{0.5}$O$_4$, while PBEsol+$U$+$V$ further enhances it, similar to the case of LiMn$_2$O$_4$. In Mn$_{1.5}$Ni$_{0.5}$O$_4$, conversely,  while PBEsol+$U$ increases the gap compared to PBEsol, PBEsol+$U$+$V$ reduces it back and falls in-between the PBEsol and PBEsol+$U$ values. Therefore, we believe that the PBEsol+$U$+$V$ band gaps of 0.76 and 0.42~eV for LiMn$_{1.5}$Ni$_{0.5}$O$_4$ and Mn$_{1.5}$Ni$_{0.5}$O$_4$, respectively, are the most reliable ones and should be used for comparisons in the future computational and experimental works.

\begin{table}[t]
\renewcommand{\arraystretch}{1.3}
\centering
\begin{tabular}{lcc}
\hline\hline
Method & \parbox{2.5cm}{\centering LiMn$_{2}$O$_4$} & \parbox{2.5cm}{\centering Mn$_{2}$O$_4$} \\ 
\hline
PBEsol                      &   0.00    &  1.05      \\
PBEsol+$U$                  &   0.73    &  1.44      \\
PBEsol+$U$+$V$              &   1.19    &  2.05      \\
PBE0r~\cite{Eckhoff:2020}&   1.1     &  2.2       \\
Expt.~\cite{Ouyang:2006} &   1.2     &            \\
\hline\hline
\end{tabular}
\caption{Band gaps (in eV) of Li$_x$Mn$_2$O$_4$ at $x=1$ and $x=0$.}
\label{tab:band_gaps_LMO}
\end{table}

\begin{table}[t]
\renewcommand{\arraystretch}{1.3}
\centering
\begin{tabular}{lcc}
\hline\hline
Method & \parbox{2.5cm}{\centering LiMn$_{1.5}$Ni$_{0.5}$O$_4$} & \parbox{2.5cm}{\centering Mn$_{1.5}$Ni$_{0.5}$O$_4$} \\ 
\hline
PBEsol                      &   0.00    &  0.37      \\
PBEsol+$U$                  &   0.33    &  0.62      \\
PBEsol+$U$+$V$              &   0.76    &  0.42      \\
\hline\hline
\end{tabular}
\caption{Band gaps (in eV) of Li$_x$Mn$_{1.5}$Ni$_{0.5}$O$_4$ at $x=1$ and $x=0$.}
\label{tab:band_gaps_LMNO}
\end{table}

\subsection{Projected density of states}
\label{sec:PDOS}

\begin{figure*}[p]
  \centering
  \includegraphics[width=0.9\linewidth]{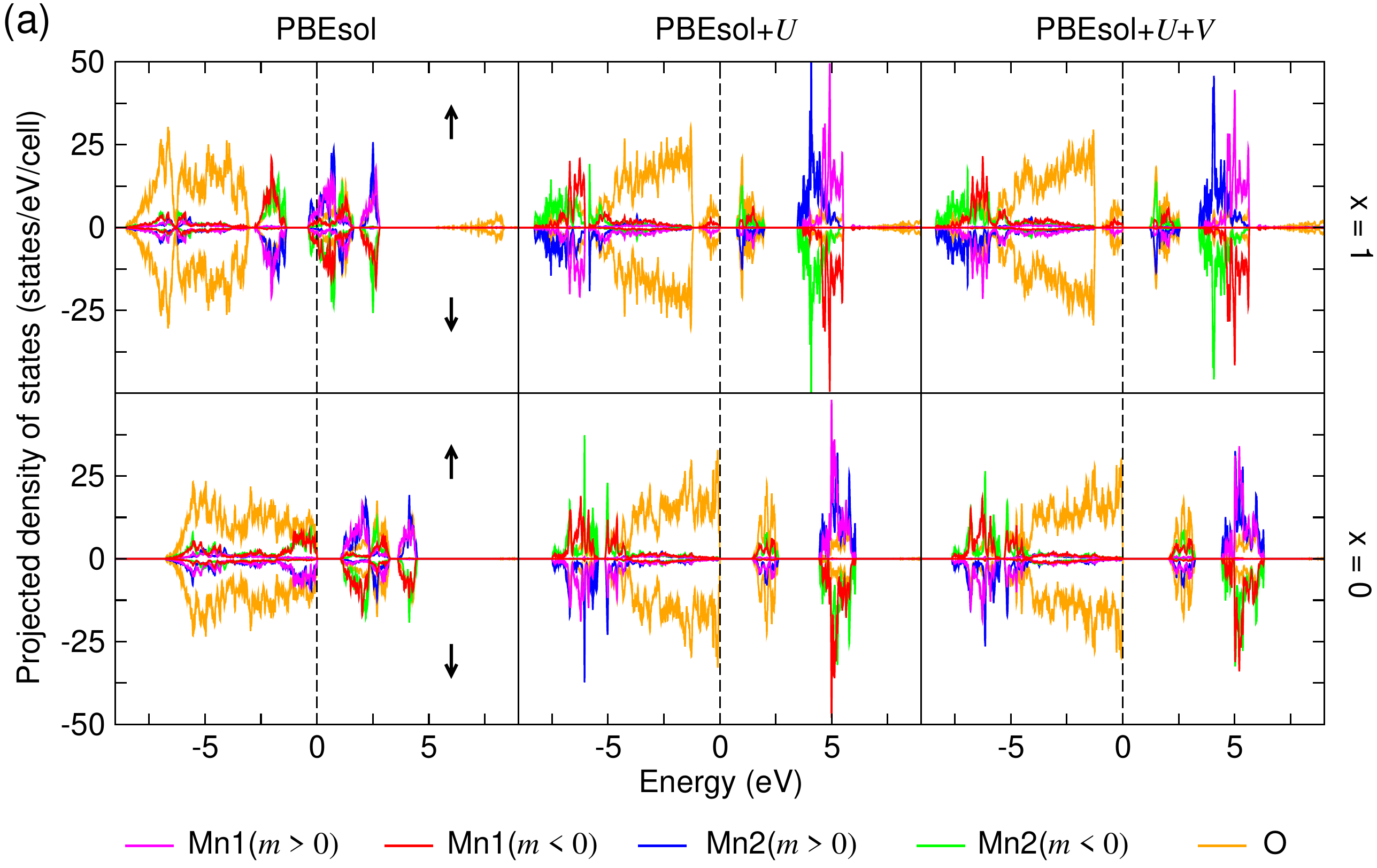}
  \vskip 0.5 cm
  \includegraphics[width=0.9\linewidth]{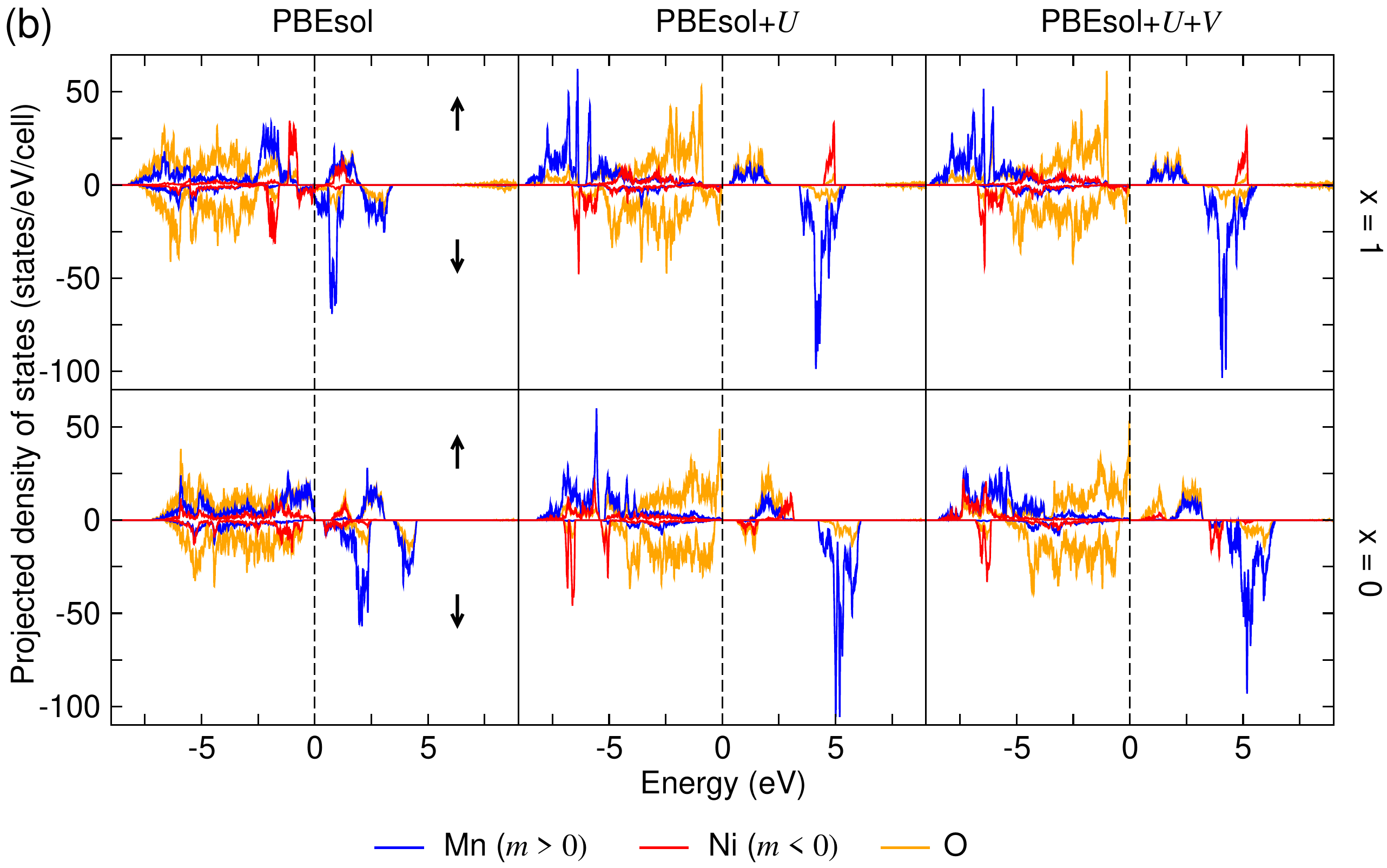}
   \caption{Spin-polarized projected density of states of (a)~Li$_x$Mn$_2$O$_4$ and (b)~Li$_x$Mn$_{1.5}$Ni$_{0.5}$O$_4$ at $x=1$ and $x=0$ computed using PBEsol, PBEsol+$U$, and PBEsol+$U$+$V$. The zero of energy corresponds to the top of the valence bands in the case of insulating ground states or the Fermi level in the case of metallic ground states. The upper and lower parts of each panel correspond to the spin-up and spin-down channels, respectively. The sign of the magnetic moment $m$ is highlighted for TM elements.}
\label{fig:PDOS}
\end{figure*}

Spin-resolved projected density of states (PDOS) for all materials studied here computed using PBEsol, PBEsol+$U$, and PBEsol+$U$+$V$ are shown in Fig.~\ref{fig:PDOS}. As was pointed out in this and previous theoretical works~\cite{Zhou:2004b, Ouyang:2009, Xu:2010, Karim:2013, Hoang:2014, Liu:2017, Isaacs:2020}, PBEsol is unable to distinguish between Mn$^{3+}$ and Mn$^{4+}$ ions in LiMn$_2$O$_4$ due to self-interaction errors leading to the overdelocalization of Mn-$3d$ states. This can be seen in Fig.~\ref{fig:PDOS}(a), where both types of Mn ions show nearly identical PDOS across the entire energy range, and a metallic solution is found. Adding the Hubbard $U$ correction to the Mn-$3d$ states allows them to localize and a clear energy separation between these Mn1-$3d$ (Mn$^{3+}$) and Mn2-$3d$ (Mn$^{4+}$) states can be seen, resulting in an insulating ground state. Adding the Hubbard $V$ correction on top only leads to minor changes in the PDOS, consistently with the findings for other TM oxides~\cite{Mahajan:2021, Mahajan:2022}. Overall, the Hubbard $U$ and $V$ corrections lead to a PDOS where the valence band maximum (VBM) is dominated by O-$2p$ states that are slightly hybridized with Mn$^{3+}$-$3d$ states, while the conduction band minimum (CBM) consists of strongly mixed O-$2p$ and Mn$^{4+}$-$3d$ states. Overall, the computed PDOS agrees well with the one from the previous PBEsol+$U$ works~\cite{Xu:2010, Ouyang:2009, Liu:2017}. In Mn$_2$O$_4$, the PDOS for Mn1-$3d$ and Mn2-$3d$ states are nearly identical as both correspond to Mn$^{4+}$. Within PBEsol, the character of the VBM shows a strong mixing of O-$2p$ and Mn$^{4+}$-$3d$ states and the character of the CBM is predominantly Mn$^{4+}$-$3d$. In contrast, within PBEsol+$U$ and PBEsol+$U$+$V$ the character of the VBM is purely O-$2p$, while that of the CBM is predominantly O-$2p$ with a rather strong mixing with the Mn$^{4+}$-$3d$ states. Therefore, we find that the inclusion of the Hubbard corrections dramatically changes the PDOS of these two compounds.

In the case of LiMn$_{1.5}$Ni$_{0.5}$O$_4$ within PBEsol, both Mn$^{4+}$-$3d$ and Ni$^{2+}$-$3d$ states cross the Fermi level and hence are responsible for the metallicity of this material, thus contradicting to experiments. The addition of the Hubbard $U$ and $V$ corrections opens a gap: the character of the VBM is O-$2p$ while that of the CBM is strongly mixed between O-$2p$ and Mn$^{4+}$-$3d$ states. In contrast, in Mn$_{1.5}$Ni$_{0.5}$O$_4$ PBEsol opens a gap, with the character of the VBM being a mixture of Mn$^{4+}$-$3d$ and O-$2p$ states, while that of the CBM is a mixture of Mn$^{4+}$-$3d$ and Ni$^{4+}$-$3d$ states. Interestingly, PBEsol+$U$ and PBEsol+$U$+$V$ give somewhat different PDOS for Mn$_{1.5}$Ni$_{0.5}$O$_4$, which is mainly due to the Ni$^{4+}$-$3d$ states. While in both cases the character of the VBM is purely O-$2p$, that of the CBM is O-$2p$ mixed with the Ni$^{4+}$-$3d$ states and both appearing either in the spin-down (PBEsol+$U$) or spin-up channel (PBEsol+$U$+$V$). As was pointed out in Sec.~\ref{sec:Oxidation_state}, this is driven by the sensitivity of the Ni$^{4+}$-$3d$ electron distribution in the spin-up and spin-down channels to the Hubbard corrections. In order to gain more insight into the PDOS for Mn$_{1.5}$Ni$_{0.5}$O$_4$, it would be desirable to perform calculations using e.g. hybrid functionals or other methods of similar accuracy, ideally accompanied with photoemission experiments.

\subsection{Voltages}

The topotactic Li-intercalation voltages can be computed using the fundamental thermodynamic definition~\cite{Aydinol:1997, Cococcioni:2019, Timrov:2022c}:
\begin{equation}
    \Phi = - \frac{E(\mathrm{Li}_{x_2}\mathrm{S}) - E(\mathrm{Li}_{x_1}\mathrm{S}) - (x_2 - x_1)E(\mathrm{Li})}{(x_2 - x_1) e} ,
    \label{eq:voltage}
\end{equation}
where S is introduced for the sake of shorthand notation and it denotes Mn$_2$O$_4$ for Li$_x$Mn$_2$O$_4$, and Mn$_{1.5}$Ni$_{0.5}$O$_4$ for Li$_x$Mn$_{1.5}$Ni$_{0.5}$O$_4$. Here, $\Phi$ is the voltage, $e$ is the electronic charge, $x_1$ and $x_2$ are the concentrations of Li, and $E$ is the total energy per formula unit. It is important to remark that $E(\mathrm{Li})$ is the total energy of bulk Li computed at the level of standard DFT (PBEsol functional) while $E(\mathrm{Li}_{x_1}\mathrm{S})$ and $E(\mathrm{Li}_{x_2}\mathrm{S})$ are computed using three functionals considered in this work: PBEsol, PBEsol+$U$, and PBEsol+$U$+$V$ ($U$ and $V$ are computed self-consistently individually for each structure)~\cite{Cococcioni:2019, Timrov:2022c}. If there are several plateaus in the voltage profile as a function of the Li concentration, it is possible to compute these by selecting corresponding $x_1$ and $x_2$ values in Eq.~\eqref{eq:voltage}. Here we compute the average voltage in the whole range of Li concentrations and thus take $x_1=0$ and $x_2=1$. We also remark that the voltage can be computed at a low and high state of charge~\cite{Bolle:2020}. It is worth noting that the entropic and pressure-volume effects are neglected since these are not significant when computing the average Li-intercalation voltages~\cite{Aydinol:1997b}.

\begin{figure}[t]
  \centering 
  \includegraphics[width=0.85\linewidth]{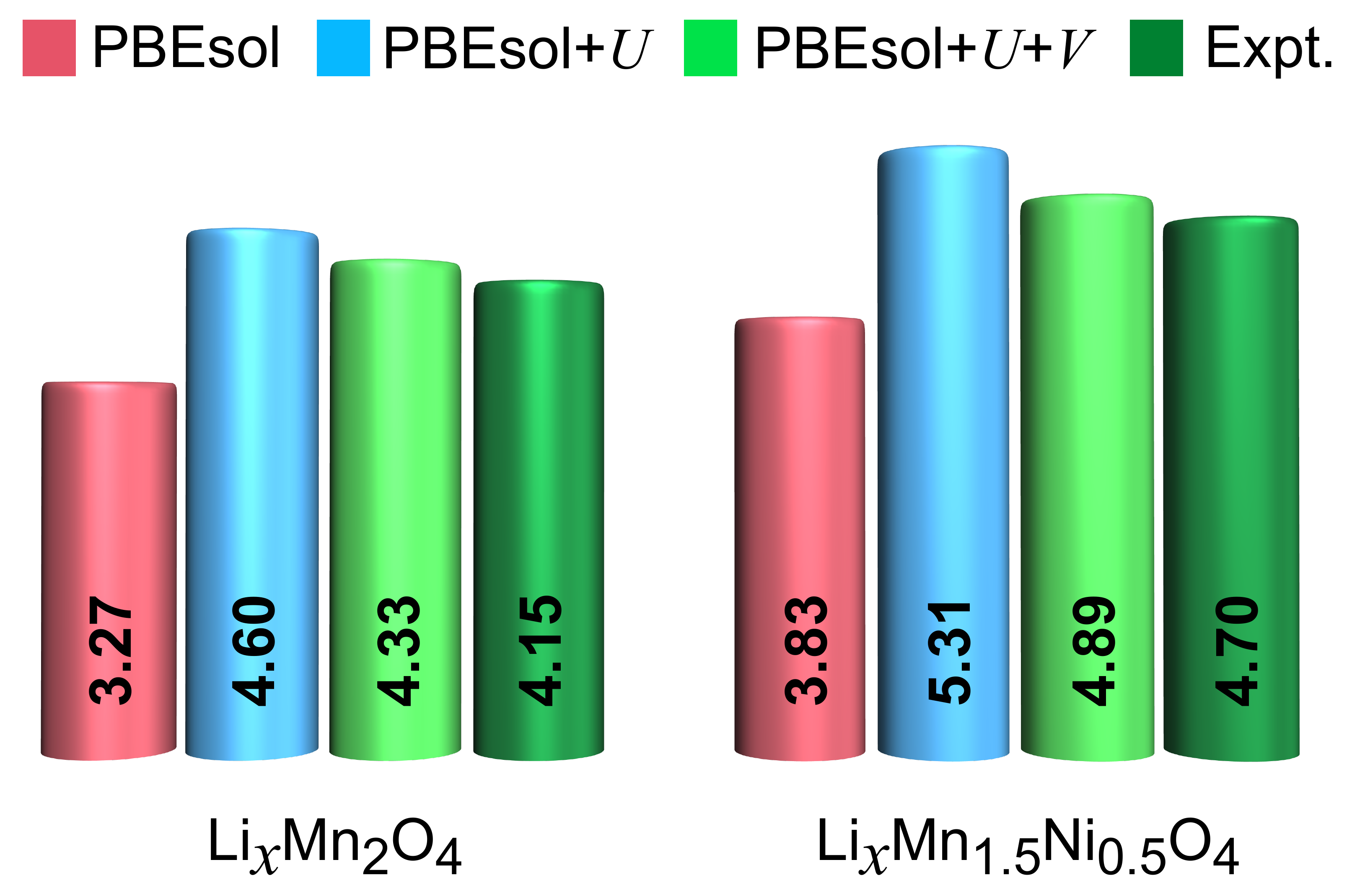}
  \caption{Voltages versus Li/Li$^+$ (in V) for Li$_x$Mn$_2$O$_4$ and Li$_x$Mn$_{1.5}$Ni$_{0.5}$O$_4$ for $0<x<1$ computed using PBEsol, PBEsol+$U$, and PBEsol+$U$+$V$. The experimental data is from Refs.~\cite{Ohzuku:1994, Barker:1995, Zhong:1997}.}
\label{fig:voltages}
\end{figure}

Figure~\ref{fig:voltages} shows the comparison of average voltages versus Li/Li$^+$ for $0<x<1$ computed using PBEsol, PBEsol+$U$, and PBEsol+$U$+$V$ and compared with experimental ones from Refs.~\cite{Ohzuku:1994, Barker:1995, Zhong:1997} (see also Ref.~\cite{Zhou:2004b}). The observed trends are the same as for phospho-olivines~\cite{Cococcioni:2019, Timrov:2022c}. Namely, PBEsol underestimates the voltages by about 20\%, while PBEsol+$U$ overestimates them by about 12\%, and PBEsol+$U$+$V$ provides the most accurate prediction of the voltages with an overestimation of about 4\% ($\sim 0.2$~V) compared to experiments. We want to stress that this is a remarkable result since here we used a fully first-principles framework with Hubbard parameters being computed from linear response and not empirically calibrated as is commonly done. The remaining discrepancies between the PBEsol+$U$+$V$ and experimental voltages could be attributed in part to finite-temperature effects that have been neglected in this study, but further studies in this direction are needed.

\section{Conclusions}
\label{sec:Conclusions}

We have presented a fully first-principles investigation of the structural, electronic, magnetic, and electrochemical properties of two prototypical spinel cathode materials -- Li$_x$Mn$_2$O$_4$ and Li$_x$Mn$_{1.5}$Ni$_{0.5}$O$_4$ ($x=0$ and $x=1$) -- using PBEsol, PBEsol+$U$, and PBEsol+$U$+$V$. The on-site $U$ and inter-site $V$ Hubbard parameters are computed self-consistently fully from first-principles using density-functional perturbation theory, hence avoiding any empiricism. We have shown that while the on-site $U$ correction is crucial to reproduce the correct trends in the electronic-structure properties of these materials, the inter-site $V$ is decisive for quantitatively accurate predictions of these. In particular, the Li-intercalation voltages are most accurate at the PBEsol+$U$+$V$ level, with a deviation of about 4\% ($\sim 0.2$~V) from experiments. Lattice parameters, cell volumes, and band gaps are more accurately predicted within PBEsol+$U$+$V$ than PBEsol+$U$, underlining again the importance of inter-site Hubbard corrections due to significant metal-ligand hybridization in these materials. We have also shown that L\"owdin occupations and magnetic moments are not reliable proxies for determining the oxidation state of transition-metal ions, while the approach of Sit and coworkers~\cite{Sit:2011} is very robust in spinel materials, in analogy with phospho-olivines~\cite{Timrov:2022c}. This allowed us to correctly identify the presence of Mn$^{3+}$ and Mn$^{4+}$ ions in LiMn$_2$O$_4$, Mn$^{4+}$ in Mn$_2$O$_4$, Mn$^{4+}$ and Ni$^{2+}$ in LiMn$_{1.5}$Ni$_{0.5}$O$_4$, and Mn$^{4+}$ and Ni$^{4+}$ in Mn$_{1.5}$Ni$_{0.5}$O$_4$. This work, therefore, paves the way for accurate future studies of other families of Li-ion cathode materials using extended DFT+$U$+$V$ functionals.

\section*{Acknowledgements}

This research was supported by the NCCR MARVEL, a National Centre of Competence in Research, funded by the Swiss National Science Foundation (grant number 205602). For the purpose of Open Access, a CC BY public copyright licence is applied to any Author Accepted Manuscript (AAM) version arising from this submission. Computer time was provided by the Swiss National Supercomputing Centre (CSCS) under project No.~s1073.

\section*{Author Contributions}

Author contributions include conceptualization, I.T. and N.M.; methodology, I.T. and N.M.;
software, I.T.; validation, I.T.; formal analysis, I.T., M.K., N.M.;
investigation, I.T., M.K., N.M.; resources, I.T. and N.M.; data curation, I.T., M.K., N.M.; writing–original draft preparation, I.T.; writing–review and editing, I.T., M.K., N.M.; visualization, I.T. and M.K.; supervision, N.M.; project administration, I.T. and N.M.; funding acquisition, N.M. All authors have read and agreed to the published version of the manuscript.

\section*{Conflicts of interest}

There are no conflicts to declare.


\begin{thebibliography}{100}
\bibitem [{Not()}]{Note_PBEsol}%
  \BibitemOpen
  \href@noop {} {}\bibinfo {note} {{To be more specific, in the following we
  use the acronym PBEsol+$U$ in place of DFT+$U$, and PBEsol+$U$+$V$ in place
  of DFT+$U$+$V$.}}\BibitemShut {Stop}%
\end{thebibliography}

%

\end{document}